\documentstyle[aps,floats,preprint,rotating,epsfig]{revtex}
\tightenlines

\begin{document}
\draft
\preprint{\vbox{%\hbox{OCIP/C 00-01}
\hbox{DESY-02-034}
\hbox{hep-ph/0207240}
\hbox{July 2002}}}
\title{Discovery Potential for Doubly Charged Higgs Bosons in $e^+e^-$ 
Collisions at LEP}
\author{Stephen Godfrey$^{a,b}$, Pat Kalyniak$^{a}$ and Nikolai Romanenko$^{a}$}
\address{
$^{a}$ Ottawa-Carleton Institute for Physics \\
Department of Physics, Carleton University, Ottawa, Canada K1S 5B6\\
$^{b}$ DESY, Deutsches Elektronen-Synchrotron, D22603 Hamburg, Germany}

\maketitle

\begin{abstract}
We study the discovery limits for doubly charged Higgs bosons,
$\Delta^{--}$, obtainable at the LEP $e^+e^-$ collider.
We expect that the LEP2 collaborations can rule out the existence of a 
doubly charged Higgs boson of mass below about 190~GeV
for Yukawa couplings greater than 0.1.  
However, even  for larger values of $M_\Delta$, 
evidence for the $\Delta$ can be seen due to contributions from 
virtual intermediate $\Delta$'s 
provided they have relatively large values of the Yukawa couplings.
\end{abstract}

\section{Introduction}

The real nature of the Higgs sector is unknown, with the single $SU(2)_L$
doublet of the Standard Model (SM) as only the simplest
possibility.  Scalar $SU(2)$ triplet representations arise in some
extensions of the SM, often motivated by their usefulness in generating
neutrino masses.  A Higgs triplet with the appropriate quantum numbers
includes a doubly charged Higgs boson, $\Delta^{--}$, as part of its
physical particle content.  The $\Delta^{--}$ has a distinct experimental
signature through its decay to a pair of like sign leptons.
The discovery of a $\Delta^{--}$ would have important 
implications for our understanding of the Higgs sector and more 
importantly, for what lies beyond the standard model. This is 
especially interesting at this time given the 
evidence for neutrino mass from SNO \cite{SNO} 
and Super-Kamiokande \cite{superK}
and the recent hints of 
dileptons by the H1 Collaboration \cite{h1} at HERA.

We have previously considered signals for doubly charged Higgs boson
production via the process $e^- \gamma \to e^+\mu^-\mu^-$
\cite{godfrey02}, where the
photon is produced with a laser backscattered from one of the beams of an
$e^+e^-$ collider. In that work we considered collider centre of mass
energies of $\sqrt{s}=500$~GeV and higher.
We refer the interested reader to ref. \cite{gregores,earlier} 
for earlier work on this process.
 Here we study the sensitivity
to a doubly charged Higgs boson in the 
process $e^+e^- \to e^+ e^+\mu^-\mu^-$  for centre 
of mass energies appropriate to LEP200, $\sqrt{s}\sim 200$~GeV.  
We approximate this reaction using the
subprocess $e^+ e^- + \gamma  \to e^+ e^+\mu^-\mu^-$, assuming
the Weizs\"acker-Williams photon spectrum \cite{ww} 
for the intermediate photon. Thus, this work involves the same
$e\gamma$ initiated process
as we have  previously considered, as shown in the Feynman diagrams of
Fig. 1. The direct production of doubly charged Higgs bosons is
illustrated by the diagrams of Fig. 1a. However, the non-resonant
contributions of the Feynman diagrams of Fig. 1b also contribute to the
distinct $\mu^-\mu^-$ signal when the mass of the doubly charge Higgs
exceeds the $e\gamma$ centre of mass energy,
$M_{\Delta}>\sqrt{s_{e\gamma}}$. These non-resonant contributions play
an important role in the reach that one can obtain for doubly charged
Higgs masses.
Because the signature of same sign 
muon pairs in the final state is so distinctive and has no SM background, 
we find that the 
process can be sensitive to virtual $\Delta^{--}$'s with masses in
excess of
the centre of mass energy, depending on the strength of the
Yukawa coupling to leptons.  

We briefly summarize the relevant properties of the doubly charged 
Higgs boson. We 
refer the reader to ref. \cite{godfrey02} for more 
details of models that include Higgs triplets such as the Left-Right
symmetric model \cite{LR}
and the Left-handed Higgs triplet model \cite{Gelmini}, and for a more
comprehensive list of references.

The generic form of the Yukawa coupling to leptons of a scalar which
transforms as an $SU(2)_L$ triplet is
\begin{equation} 
\label{eq1}
{\cal L}_{Yuk} =- i h_{ll'} \Psi_{l L}^T C \sigma_2 \Delta\Psi_{l' L}+{
h.c.}, 
\end{equation}
where $\Delta$ denotes the scalar triplet, 
 $ \Psi_{l L}$ denotes the left-handed lepton doublet with flavour $l$,
and $C$ is the charge conjugation matrix.  
Thus the interaction of the
doubly charged Higgs boson with leptons is parametrized by a set of
couplings $h_{ll'}$, where $l,l'$ are flavour indices.

The product of Yukawa couplings $h_{ee}h_{\mu\mu}$ dictates the 
magnitude of the process we consider here.
 Indirect
constraints on the masses and couplings of doubly charged Higgs bosons
have been obtained from lepton 
number violating processes and muonium-antimuonium conversion experiments 
\cite{swartz,hm,fujii,chk}.  The most stringent limits come from that latter 
measurement \cite{swartz,muoniumex}.  For flavour diagonal couplings these 
measurements 
require that the ratio of the Yukawa coupling, $h$, and Higgs mass, 
$M_\Delta$, satisfy
$h/M_\Delta < 0.44$~TeV$^{-1}$ at 90\% C.L..
Non-diagonal flavour couplings are indirectly constrained to be rather 
small by the
rare 
decays $\mu\rightarrow \bar{e} ee$ \cite{mu3e,pdb} and
$\mu\rightarrow e\gamma$ \cite{mueg} so we neglect them in this 
analysis.

The indirect bounds allow the existence of low-mass doubly charged Higgs with 
a small coupling constant.  These limits can be 
circumvented in certain models as a result of cancellations 
among additional diagrams arising from other new physics 
\cite{framras,pleitez}.  Thus, 
from this point of view, direct limits are generally more robust.  We
refer the reader to ref. \cite{godfrey02} for references on other direct
bounds.  We only specifically note here the work of
Gregores {\it et al}. \cite{gregores}, 
which most closely resembles the approach presented here.  We point out 
certain differences between our calculation and theirs in the next
section.

\section{Calculations and Results}

We are studying the sensitivity to
doubly charged Higgs bosons in the process 
$e^+e^- \to e^+ e^-\gamma \to e^+ e^+\mu^-\mu^-$ where the intermediate 
photon is given by the Weizs\"acker-Williams photon spectrum \cite{ww}.
 The signal of like-sign dimuons is distinct and SM background free, 
offering excellent potential for $\Delta^{--}$ discovery. 
The direct production of doubly charged Higgs bosons and their 
non-resonant contributions are illustrated in Figs. 1a and 1b,
respectively.

To calculate the cross section we convolute the Weizs\"acker-Williams 
photon spectrum, $f_{\gamma/e}(x)$ \cite{ww}, with the subprocess 
cross section, $\hat{\sigma}(e^- \gamma \to e^+ \mu^-\mu^-)$;
\begin{equation}
\sigma= \int dx  f_{\gamma/e}(x,\sqrt{s}/2) \;
\hat{\sigma}(e^- \gamma \to e^+ \mu^-\mu^-).
\end{equation}
We obtained analytic expressions for the 
matrix elements using the CALKUL helicity amplitude 
method \cite{CALCUL}
and performed the phase space integrals using Monte-Carlo integration
techniques.  This 
approach offers a nice check using the gauge invariance 
properties of the sum of the amplitudes.  As the expressions for the 
matrix elements  are lengthy and not particularly illuminating we do 
not include them here.  Our numerical results include the possibility of
both charge states, where the photon is emitted from either
the electron or the positron.

Because we are including contributions to the final state that proceed 
via off-shell $\Delta^{--}$'s we must include the doubly-charged Higgs 
boson width in the $\Delta^{--}$ propagator.  The $\Delta$ width
is dependent on the parameters of the model, 
which determine the size and relative importance of various
decay modes.
The decay $\Delta^{--} \to W^- W^-$ is negligible assuming
the VEV of the neutral member of the triplet is small, which it is
limited to be by the LEP2 bound on the $\rho$ parameter \cite{LEP}
 and by
 bounds on the mass of left-handed neutrinos \cite{numass,SNO}.  
The decay rates for $\Delta^{--} \to \Delta^- W^-$
and $\Delta^{--} \to \Delta^- \Delta^- $
depend both on the model's couplings and on the Higgs mass 
spectrum, the latter consideration determining 
whether the decays are kinematically allowed at all.
For the $\Delta^{--}$ mass range we are 
considering here we 
expect the partial width to final state fermions, which is given by
\begin{equation}
\Gamma (\Delta^{--}\to \ell^- \ell^-) = \frac{1}{8\pi} 
h^2_{ \ell \ell} M_\Delta,
\end{equation}
to dominate.  
Since we assume in our numerical results that $h_{ ee} =h_{\mu\mu}
=h_{\tau\tau} \equiv h$,
we have 
$\Gamma_f = 3 \times \Gamma (\Delta^{--}\to \ell^- \ell^-) $.
However, to be general, we also allow the possibility of a small 
additional width  by taking
\begin{equation}
\Gamma (\Delta^{--}) = \Gamma_b + \Gamma_f,
\end{equation}
where $\Gamma_b$ is the partial width to final state bosons and 
$\Gamma_f$ is the partial width into final state SM fermions.
Indeed, the $\Delta$ width cannot be given without 
knowing the full particle content of the theory 
so that $\Gamma_b$ can be regarded as reflecting this ignorance.
We consider 
two scenarios for the $\Delta$ width: a scenario with 
$\Gamma_b=0$ and a scenario with $\Gamma_b=1.5$~GeV for $M_\Delta>160$~GeV, 
a possible threshold for boson pair production.  In fact, the non-zero 
bosonic width only makes a noticeable contribution to our results 
for $M_\Delta$ just above this threshold value and we include it only 
to give the reader an idea of the uncertainty due to our ignorance of 
$\Gamma_b$.  

A final note before proceeding to our results is that we only consider 
chiral couplings of the $\Delta$ to leptons.  Our results here are all
based on left-handed couplings.  However,  we also did the 
calculations for right-handed couplings.  The amplitude squared and 
hence the numerical results are identical in both cases.  As a result, 
the discovery potential for $\Delta_R$ would be the same as that 
for $\Delta_L$, assuming one can make 
the same assumptions regarding parameters and mass spectra. This 
assumption may not be valid.  We did not consider 
the case of mixed chirality.

To obtain numerical results we take as the SM inputs
$\sin^2\theta_W = 0.23124$ and $\alpha=1/128$
\cite{pdb}.  Since we work only to leading order in $|{\cal M}|^2$,
there is some arbitrariness in what to use for the above input,
in particular $\sin^2\theta_W$.

We consider two possibilities for the $\Delta^{--}$ signal.  In the 
first case we impose that the three subprocess final state particles be 
observed and identified.  
This has the advantage that the event can be fully 
reconstructed and as a check, the momentum must be balanced, at least 
in the transverse plane.  In the second case, we assume that the
positron is not observed, having been lost down the beam pipe.  
This case has the 
advantage that the cross section is enhanced due to divergences 
in the limit of massless fermions.  The disadvantage is 
that there will be some missing energy in 
the reaction so that it cannot be kinematically 
constrained which might lead to backgrounds where some particles in SM 
reactions are 
lost down the beam.  Although we expect these potential backgrounds
to be minimal, this issue needs to be understood in a realistic 
detector simulation and should be kept in mind.

To take into account detector acceptance we restrict 
the angles of the observed particles relative to the beam, 
$\theta_{\mu},\; \theta_{e^+}$, to the ranges $|\cos \theta| \leq 
0.95$ \cite{rob}.
We assume an identification efficiency for each of the
detected final state particles of $\epsilon = 0.9$.  
Finally, we note that one could impose additional 
cuts on the lepton energies to reflect further details of detector 
acceptance and impose a maximum 
value on the muon energies so that the tracks are not so stiff that 
their charge cannot be determined.  However, these details 
are best left to the experimentalists to decide in the context of 
their detector simulation.

In Fig. 2 we show the cross sections as a function of $M_\Delta$ 
for the reaction $e^+e^- \to e^+ e^+ \mu^-\mu^- +
e^- e^- \mu^+\mu^+ $ at $\sqrt{s}=200$~GeV
for the two final state situations.  
The value of the Yukawa coupling is taken arbitrarily to be $h=0.1$ and 
we show cross sections for the two treatments of the $\Delta^{--}$ width
described above. 
The  solid (dotted) curve is for the case when
all three subprocess final state leptons are observed in the detector and 
$\Gamma_\Delta=\Gamma_f$ ($\Gamma_\Delta=\Gamma_f+\Gamma_b$) and the
dashed and dot-dashed curves arise when only the final state muons are 
observed for those two width scenarios.  The cross sections fall
monotonically as $M_{\Delta}$ increases, with a steep dropoff beyond the
$\Delta$ production threshold, $M_\Delta > \sqrt{s}$, to rather small
values.
The cross section for the case of only observing the final 
state muons is similar in shape although 
roughly a factor of 2 larger in magnitude.

Although the cross sections below 
threshold are rather small, the signature of a $\Delta^{--}$ 
is so distinctive and 
SM background free, discovery would be signalled by even one event.
With the LEP integrated luminosities we expect that 
$M_{\Delta}$ can probed up to 
the kinematic limit of $\sim 190$~GeV for moderate values 
of $\Delta -ff$ Yukawa couplings but a $\Delta^{--}$ with larger mass
would reveal itself if its Yukawa coupling were large enough.
To obtain discovery limits for the 
$\Delta^{--}$ we use the integrated luminosities and corresponding 
centre of mass energies given in Table I. To generate 
results 
based on the total integrated luminosity for all four LEP experiments, we
multiplied the integrated luminosities per experiment given in Table I
by a factor of four.  
To obtain discovery limit contours we calculated the
expected event rate for each value of $h$ and $M_\Delta$.  This was 
done by calculating the cross section for 
each value of $h$ and $M_\Delta$ for a given $\sqrt{s}$ and 
multiplying  
the cross section by the detection efficiency and  
integrated luminosity corresponding to that $\sqrt{s}$.  
The event rates corresponding to  each value of 
$\sqrt{s}$ were then added together to give the total expected event rate for 
each value of $h$ and $M_\Delta$.

In Fig. 3 we show 95\% C.L. contours for the discovery of a doubly 
charged Higgs boson, corresponding to 3 events, in the 
Yukawa coupling - doubly charged Higgs mass ($h-M_{\Delta}$) 
parameter space.  The four contours correspond to the case of 
observing the final state $\mu$ pairs plus the beam positron (or 
electron) with $\Gamma_\Delta = \Gamma_f$ (long dashed line) and 
$\Gamma_\Delta = \Gamma_f +\Gamma_b$ (dotted line) and the cases of 
observing only the $\mu$ pair 
with $\Gamma_\Delta = \Gamma_f$ (dot-dashed line) and 
$\Gamma_\Delta = \Gamma_f +\Gamma_b$ (short-dashed line).

One sees that the discovery limits are only sensitive to the $\Delta$ 
width for a small range of $M_\Delta$ when $\Gamma_b$ is ``turned 
on'': $160 \; \hbox{GeV} < M_\Delta < 200 \;\hbox{GeV}$.  

It should be noted in comparing our results to 
those of Gregores {\it et al} \cite{gregores}, that they
only included the
Feynman diagrams with s-channel contributions from the $\Delta^{--}$
and therefore restricted their study to resonance $\Delta^{--}$ production.
In effect they looked at $e^-e^-$ fusion where one of the $e^-$'s is 
the beam electron and the other arises from the equivalent 
particle approximation for an electron in the photon.  In this 
approximation the authors assume that the positron is lost down the 
beam.  Furthermore,
they have not included the bosonic width. This is quite typical for 
doubly charged Higgs studies; however, as we show, the results are
sensitive to this parameter for a particular range
of $\Delta$ masses. Additionally, in ref. \cite{gregores},
they take an overall efficiency factor of 0.9 while ours is 
 a more conservative $(0.9)^2$ for the signature in which only the two 
final state muons are observed. (Note also that the coupling of
ref. \cite{gregores}
is related to ours by $\lambda=h/\sqrt{2}$.) Taking into account this
width dependence and the fact that ours is 
a complete calculation
rather than an equivalent particle approximation, explains the difference
between the results. 

The general behavior of these sensitivity curves reflects the 
dependence of the cross section on $M_\Delta$.  
For $\sqrt{\hat{s}_{e\gamma}}>M_\Delta$ real $\Delta$'s
can be produced and
the process is sensitive to the existence of $\Delta$'s with 
relatively small Yukawa couplings.  
However, for $M_\Delta > \sqrt{\hat{s}_{e\gamma}}$ the process only 
proceeds via virtual $\Delta$'s resulting in a smaller cross section 
that only leads to observable signals for larger values of the Yukawa 
couplings.  We summarize the 
95\% probability mass discovery limits 
 for the various scenarios in Table II for some representative coupling
values.

\section{Summary}

Doubly charged Higgs bosons arise in one of the most straightforward 
extensions of the standard model: the introduction of Higgs triplet 
representations.  Their observation would signal physics outside the 
current paradigm and perhaps point to what lies beyond the SM.  As 
such, searches for doubly charged Higgs bosons are important.
In this paper 
we studied the discovery potential for  doubly charged 
Higgs bosons at LEP2. We found that doubly 
charged Higgs bosons could be discovered for even relatively small 
values of the Yukawa couplings: $h > 0.01$ for $M_\Delta$ up to 
roughly 190~GeV.   
We note that the OPAL Collaboration at LEP has used this process to 
infer a 95 \% C.L. upper limits of $h_{ee}<0.08$ for $M_\Delta < 
160$~GeV \cite{opal}.
For values of $M_\Delta$ greater than the 
production threshold, discovery is possible even for $M_\Delta$ 
greater than
$\sqrt{s}$ because of the distinctive, background free final 
state in the process  $e\gamma \to e^+ \mu^-\mu^-$ which can proceed 
via virtual contributions from intermediate $\Delta$'s.

\acknowledgments

The authors are most grateful to Rob Macpherson for many useful 
discussions and communications.
This research was supported in part by the Natural Sciences and Engineering 
Research Council of Canada.  N.R. is partially supported by
RFFI Grant 01-02-17152  (Russian Fund of Fundamental Investigations).

\newpage
\begin{figure}
\centerline{\epsfig{file=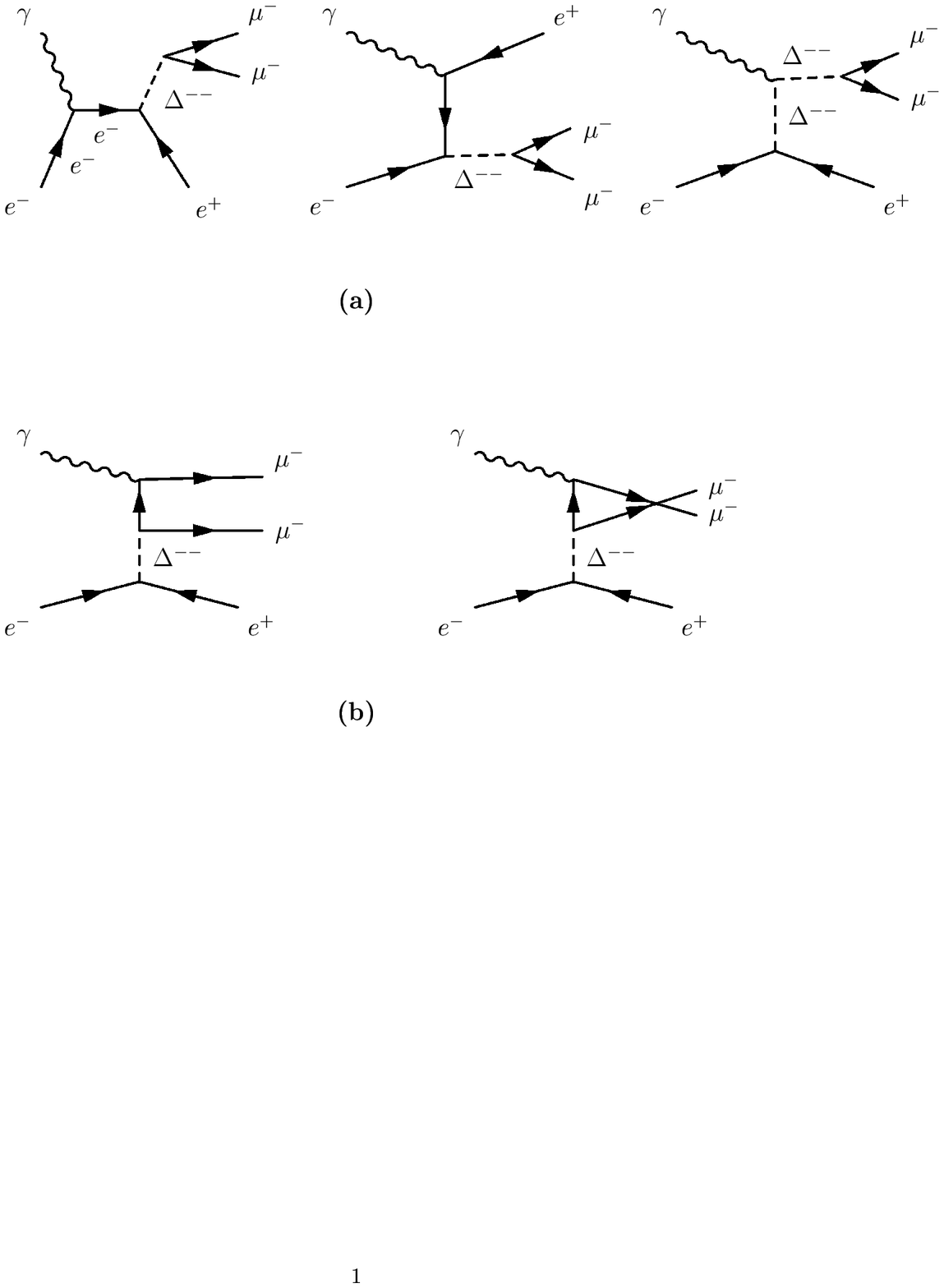,width=6.5in}}
\vspace{20pt}
\caption{The Feynman diagrams contributing to doubly charged Higgs 
boson production in $e^- \gamma \to e^+ \mu^-\mu^-$.}
\label{Fig1}
\end{figure}

\newpage
\begin{figure}
\centerline{\epsfig{file=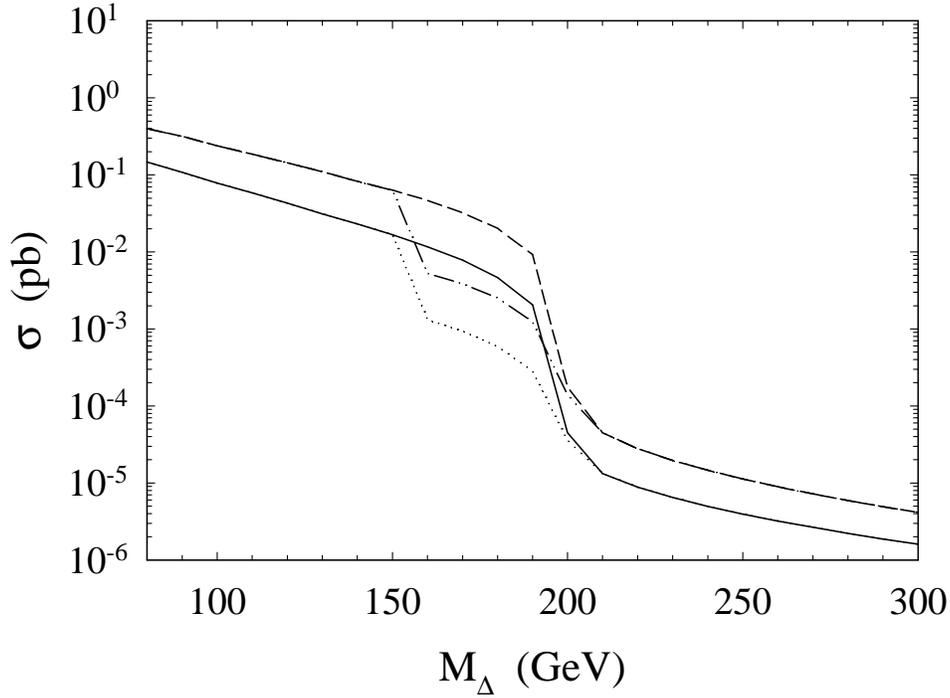,width=5.0in}}
\vspace{20pt}
\caption{The cross section, $\sigma(e^+e^- \to e^+ e^+ \mu^-\mu^-)$
as a function of $M_{\Delta}$ for $\sqrt{s}=200$~GeV and $h=0.1$. The
solid and dotted curves are for all three final state particles 
being detected using $\Gamma_{\Delta} = \Gamma_f$ and $\Gamma_{\Delta} =
\Gamma_f + \Gamma_b$, respectively.  The dashed and dot-dashed curves
correspond to the case of  
only the $\mu^-\mu^-$
pairs observed with the positron lost down the beam-pipe, for the
same two width scenarios, respectively.}
\label{Fig2}
\end{figure}

\newpage
\begin{figure}
\centerline{
\begin{turn}{-90}
\epsfig{file=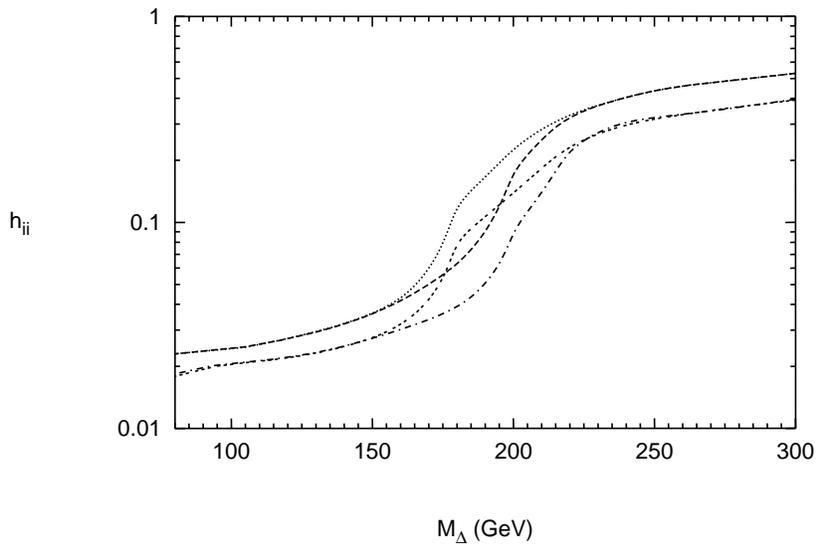,width=7.5cm,clip=}
\end{turn}
}
\vspace{20pt}
\caption{Discovery limits for doubly charged Higgs bosons as a 
function of the Yukawa coupling and $M_\Delta$ at LEP2
for different observed final states and 
two $\Gamma_\Delta$ scenarios. 
The long-dashed and dotted curves are for observing the dimuons and 
electron or positron in the final state for $\Gamma_\Delta =\Gamma_f$ 
and $\Gamma_\Delta =\Gamma_f+\Gamma_b$ respectively while the
dot-dashed and short-dashed curves are for the cases of seeing only 
the final state muons for $\Gamma_\Delta =\Gamma_f$ 
and $\Gamma_\Delta =\Gamma_f+\Gamma_b$ respectively.
The limits are based on observation of three events assuming the
integrated luminosities given in Table I.
}
\label{Fig3}
\end{figure}

\newpage

\begin{table}[t]
\caption{LEP2 centre of mass energies and integrated luminosites used 
in our analysis.
}
\label{luminosities}
\vspace{0.4cm}
\begin{center}
\begin{tabular}{rrr}
%\hline
Year & $\sqrt{s}$ & $L$/experiment \\
     & (GeV)  & ($pb^{-1}$)  \\
\hline
1997 & 	 182.7 & 55   \\
1998 & 	 188.6 & 180   \\
1999 & 	 197.4 & 221  \\
2000 & 	 206.0 & 220   \\
\end{tabular}
\end{center}
\end{table}

\begin{table}[t]
\caption{95\% probability mass discovery limits of doubly charged 
Higgs bosons, given in GeV for LEP2.  We use the centre of mass 
energies and integrated luminosities given in Table I.
The cases shown are for $e^+ \mu^-\mu^-$ detected and for the $e^+$ 
lost down the beam.
}
\label{limitstabbl}
\vspace{0.4cm}
\begin{center}
\begin{tabular}{lcccc}
%\hline
$h_{\ell\ell}$ & \multicolumn{2}{c}{$e^+\mu^-\mu^-$ observed} 
	& \multicolumn{2}{c}{$\mu^-\mu^-$ observed}  \\
	& $\Gamma_\Delta=\Gamma_f$ & $\Gamma_\Delta=\Gamma_f+ \Gamma_b$
	& $\Gamma_\Delta=\Gamma_f$ & $\Gamma_\Delta=\Gamma_f+ \Gamma_b$ \\
      & $M_{\Delta}$ (GeV)  & $M_{\Delta}$ (GeV)  
	& $M_{\Delta}$ (GeV)  & $M_{\Delta}$ (GeV) \\
\hline
%0.02  & - & - & 95 & 95   \\
0.03  & 130 & 130 & 160 & 155  \\
0.05  & 170 & 165 & 190 & 174   \\
0.1   & 192 & 178 & 202 & 187   \\
\end{tabular}
\end{center}
\end{table}

\end{document}